\documentclass[aps,twocolumn,showpacs]{revtex4}
\usepackage{amsmath}%
\usepackage{amsthm,amsmath,amssymb}
\usepackage{comment}
\usepackage{mathrsfs}
\usepackage{graphicx}% Include figure files
\usepackage{dcolumn}% Align table columns on decimal point
\usepackage{bm}% bold math
\usepackage{color, soul}
\usepackage{color, xcolor}
\usepackage[utf8]{inputenc}
\usepackage[T1]{fontenc}
\usepackage{mathptmx}
\usepackage{multirow}
\usepackage{makecell}
\usepackage{setspace}
\usepackage{footnote}
\usepackage{hyperref}
\hypersetup{hypertex=true,
	colorlinks=true,
	linkcolor=blue,
	anchorcolor=blue,
	citecolor=blue}

\begin{document}
\title{Soliton eigenvalue control by interaction of circularly polarized lights in a nonlinear fiber}
%\title{A high-efficient physical approach for soliton eigenvalue control in nonlinear fibers}
%\title{A high-efficient physical approach for decomposing and reconstructing high-order solitons}

\author{Peng Gao}\email{gaopeng@xidian.edu.cn}
\author{Xiaofang Wang}%\email{xfwang@xidian.edu.cn}
\author{Sha An}%\email{gaopeng@xidian.edu.cn}
\author{Kai Wen}%\email{kai.wen@xidian.edu.cn}
\author{Juanjuan Zheng}%\email{gaopeng@xidian.edu.cn}
\author{Tanping Li}%\email{gaopeng@xidian.edu.cn}
\author{Peng Gao}\email{peng.gao@xidian.edu.cn}

\address{School of Physics, Xidian University, Xi’an 710071, China}
%\address{$^2$School of Physics, Northwest University, Xi’an 710127, China}
%\address{$^2$Key Laboratory of Optoelectronic Perception of Complex Environment, Ministry of Education, Xi’an 710071, China}
\address{Engineering Research Center of Information Nanomaterials, Universities of Shaanxi Province, Xi’an 710071, China}
\address{Xi’an Engineering Research Center of Super-resolution Optical Microscopy, Xi’an 710071, China}
%\address{$^5$Shaanxi Key Laboratory for Theoretical Physics Frontiers, Xi’an 710127, China}

 %%%%%%%%%%%%%%%%%%%%%%%%%%%%%%%%%%%%%%%%%%%%%%%%%
%\date{Mar. 1, 2018}
\begin{abstract}
We propose a physical method for controlling soliton eigenvalues in optical fibers, which is realized through the interaction between circularly polarized lights. Using this method, we not only achieve the decomposition of high-order solitons (HOSs) with different orders, but also realize physical processes of reconstructing HOSs for the first time. Compared with existing methods, our approach ensures accurate measurement of the discrete eigenvalues of HOSs while exhibiting higher decomposition efficiency.
It is worth noting that the probe soliton, which induces these phenomena, plays a key role. The requirement for a moderate steepness of the probe suggests the presence of an uncertainty principle in the measurement of soliton eigenvalues, similar to the detection of microscopic particles.
Our results can deepen the understanding of microscopic properties of solitons and their interaction mechanisms, and moreover provide a promising all-optical solution for the design of eigenvalue-based multiplexers and demultiplexers.
\end{abstract}

\maketitle

\section{Introduction}

Optical solitons are fundamentally important phenomena arising in nonlinear optical systems \cite{Kivshar-book}.
Since predicted theoretically \cite{McCall-1967,Hasegawa-1973}, they have evolved significantly over five decades, leading to diverse physical effects and practical applications \cite{Taylor-book,Blanco-Redondo-2023}.
%Optical solitons are important phenomena unique to nonlinear optical systems. Since the concept of optical solitons was first introduced in the last century, more than fifty years have passed, during which numerous related phenomena and applications have emerged. 
Physically, a soliton typically arises from the balance mechanism between dispersion and nonlinear effects.
Mathematically, for an integrable propagation model, a fundamental soliton corresponds to an eigenstate of the nonlinear Schrödinger equation (or its analogues), with a certain Zakharov-Shabat eigenvalue determined by the associated Lax equations \cite{Zakharov-1971}; in contrast, a high-order soliton (HOS) is the nonlinear superposition of fundamental solitons and thus has several different discrete eigenvalues \cite{Agrawal-book}.
The Zakharov-Shabat eigenvalues define an important physical characteristic of steepness (or pulse energy) of solitons, which has been applied to signal processing \cite{Turitsyn-2017,Perego-2024}, optical communication \cite{Prilepsky-2014,Le-2017}, and wave generations \cite{Akhmediev-2009,Soto-Crespo-2016,He-2013,Liu-2015,Duan-2019,Liu-2024}. 
Therefore, the control of soliton eigenvalues has been a significant physical issue in nonlinear optics and soliton research.

The term "eigenvalue control" originally arose from studies of spatial optical solitons \cite{Aleshkevich-2004,Kartashov-2004}. 
For HOS beams propagating in a nonlinear medium, it is possible to induce and control their eigenvalue splitting by quadratic nonlinear effect \cite{Torres-1997}, or by designing specific refractive index distributions \cite{Suryanto-2002,Aleshkevich-2004,Kartashov-2004,Zhou-2010}. 
Through this splitting process, the HOS can be decomposed into several fundamental solitons, agreeing well with the distribution of discrete eigenvalues.
In fact, research on temporal optical solitons had already demonstrated eigenvalue splitting earlier, achieved by applying perturbations to HOS in nonlinear fibers \cite{Satsuma-1974}.
Later studies proposed other methods to achieve this process, including the presence of stimulated Raman or two-photon absorption effect \cite{Tai-1988,Silberberg-1990,Afanasjev-1995}, and the use of filtering or pre-chirping technique \cite{Friberg-1992,Krylov-1999}.
In general, these methods exert perturbations either on the incident wave or on the system model.
%the eigenvalue splitting processes induced by them can fundamentally be viewed as perturbed propagation of high-order solitons.

Notably, a novel alternative approach was proposed: eigenvalue control can be realized through collisions between a weak fundamental soliton and a HOS, along with a proposed application for eigenvalue-based communication \cite{Hasegawa-1993}.
%A similar effect was demonstrated in 2014 through the collision of two high-order solitons. 
Compared with the perturbation-based approaches, the collision-based one can generally yield more standard fundamental solitons after splitting, enabling more accurate eigenvalue measurements.
However, due to the small frequency difference between after-splitting solitons, this approach requires long propagation distances to achieve sufficient splitting, thereby implying low efficiency.
This limitation motivates us to explore methods that ensure higher efficiency while maintaining accurate measurements.

In this paper, we propose a physical method for soliton eigenvalue control in optical fibers, enabling both the decomposition and reconstruction of HOSs. 
The method relies on collisions between a left-handed circularly polarized (LCP) fundamental soliton (called probe soliton) and a right-handed circularly polarized (RCP) HOS. 
Compared to the existing approaches, this strategy achieves higher efficiency in decomposition/reconstruction processes while maintaining high-precision eigenvalue measurement. 
A critical parameter governing this process is the steepness (closely related to pulse energy) of the probe soliton.
It must be optimized within a moderate range to ensure balanced control of efficiency and accuracy, which indicates the presence of an uncertainty principle.

\begin{figure*}
	\centering
	\includegraphics[width=176mm]{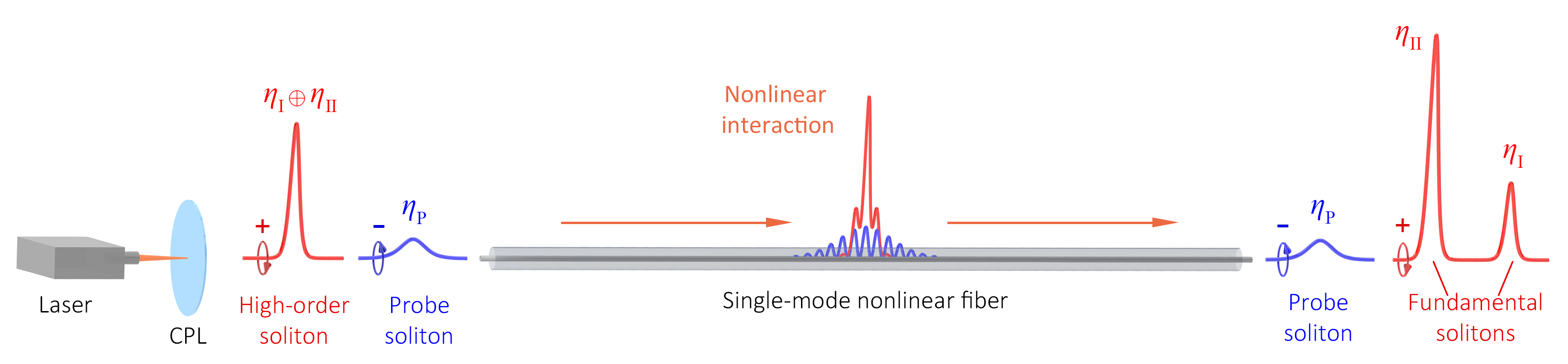}
	\caption{Schematic diagram of the process that a HOS in the RCP component ($+$) splits into two fundamental solitons
	after colliding with a fundamental soliton in the LCP component ($-$), which is called "probe soliton".
	The after-collision fundamental solitons have the eigenvalues $\eta_{\rm I}$ and $\eta_{\rm II}$, which are just the discrete eigenvalues constructing the HOS (denoted by $\eta_{\rm I}\oplus\eta_{\rm II}$).
	The CPL represents a circular polarizer.}
	\label{pic-model}
\end{figure*}

\section{Physical model}

The schematic diagram in Fig. \ref{pic-model} illustrates the physical process that we use to realize eigenvalue control, i.e., an eigenvalue splitting process of a HOS induced by the interaction of circularly polarized lights. 
%For simplicity, only the essential parts are depicted here. 
Taking the splitting of a second-order soliton as an example, we describe the general physical process as follows: First, a laser beam is converted into RCP or LCP light by a circular polarizer. 
We set the LCP component as a lower-energy fundamental soliton (i.e., a first-order soliton) to be initially injected into the fiber, while the subsequently injected RCP component is a higher-energy second-order soliton state. 
A HOS comprises two discrete eigenvalues, denoted by $\eta_{\rm I}$ and $\eta_{\rm II}$ (with $\eta_{\rm I}<\eta_{\rm II}$). 
Since the role of this fundamental soliton is only to induce the splitting of the HOS, we designate it as the "probe soliton".
To concisely represent their nonlinear superposition in this context, we adopt the symbolic notation $\eta_{\rm I}\oplus\eta_{\rm II}$ to denote the eigenvalue construction of the corresponding second-order soliton.
Here, the RCP component is deliberately assigned a slightly lower frequency than the LCP component. 
This frequency difference enables nonlinear interactions (or collisions) between the two polarized components during their propagation through the fiber. 
Following the interaction, the second-order soliton within the RCP component splits into two fundamental solitons, which exit the fiber first. 
Their eigenvalues are separately $\eta_{\rm I}$ and $\eta_{\rm II}$, which approximatively match the discrete eigenvalues constituting the original second-order soliton. 
Subsequently, the probe soliton (in LCP component) exits the fiber, exhibiting only minor changes in shape and velocity.

To study the propagation of a polarized light in a single-mode nonlinear fiber, we write the vector of electric field intensity as the form of $\vec{E}=\frac{1}{2}[\vec{A}\,e^{i(\beta z-\omega t)}+\vec{A}^*e^{-i(\beta z-\omega t)}]$,
where $\vec{A}$ and $e^{i(\beta z-\omega t)}$ are separately the slowly and fast varying parts.
In the laboratory coordinate system, considering that the propagation direction is along the $z$ axis, the slowly varying complex amplitude $\vec{A}$ can be decomposed into components in the $x$ and $y$ directions by $\vec{A}=A_x\hat{e}_x+A_y\hat{e}_y$, where $\hat{e}_x$ and $\hat{e}_y$ are separately the basis vectors in the $x$ and $y$ directions.
After substituting $\vec{E}$ into Maxwell equations and making $|A|^2$ represent the optical power by introducing a trivial factor, one can obtain the propagation equations of $A_x$ and $A_y$ \cite{Agrawal-book},
\begin{align}\label{eq-model-lp}
	&i\partial_zA_x-\frac{\beta_2}{2}\partial_t^2A_x+\sigma\Big[(|A_x|^2+\frac{2}{3}|A_y|^2)A_x+\frac{1}{3}A_x^*A_y^2\Big]\!=0,\nonumber\\
	&i\partial_zA_y-\frac{\beta_2}{2}\partial_t^2A_y+\sigma\Big[(\frac{2}{3}|A_x|^2+|A_y|^2)A_y+\frac{1}{3}A_x^2A_y^*\Big]\!=0,
\end{align}
where $\beta_2$ and $\sigma$ are the coefficients of group-velocity dispersion and Kerr nonlinear effect, respectively.
In the model (\ref{eq-model-lp}), the retarded time $t$ is introduced to construct a coordinate system that moves with the center of the optical envelope.

The vector $\vec{A}$ can also be decomposed into circularly polarized components by $\vec{A}=A_+\hat{e}_++A_-\hat{e}_-$, where $\hat{e}_+$ and $\hat{e}_-$ are separately the RCP and LCP basis vectors.
The circularly polarized components can be obtained by the transformation 
\begin{align}\label{eq-cplp}
	\begin{bmatrix}
		A_+\\A_-
	\end{bmatrix}=\frac{1}{\sqrt{2}}\begin{bmatrix}
	1 & i\\ i & 1
	\end{bmatrix}\begin{bmatrix}
	A_x\\A_y
	\end{bmatrix}.
\end{align}
Then, one can obtain the propagation equations of circularly polarized components \cite{Agrawal-book},
\begin{align}\label{eq-model-cp}
	&i\partial_zA_+-\frac{\beta_2}{2}\partial_t^2A_++\frac{2}{3}\sigma(|A_+|^2+2|A_-|^2)A_+=0,\nonumber\\
	&i\partial_zA_--\frac{\beta_2}{2}\partial_t^2A_-+\frac{2}{3}\sigma(2|A_+|^2+|A_-|^2)A_-=0.
\end{align}
The model (\ref{eq-model-cp}) describes the propagation of RCP and LCP lights in a nonlinear fiber, without considering high-order dispersion or nonlinear effects.
It is more convenient to study the interaction between circularly polarized components than the model (\ref{eq-model-lp}).

By changing the form of transformation (\ref{eq-cplp}), one can also obtain the propagation equations of elliptically polarized states (where the ratio of cross-phase modulation coefficient relative to self-phase one is between $\frac{2}{3}$ and $2$) \cite{Agrawal-book}.
In vector propagation models, some studies have revealed interaction mechanism between orthogonally polarized solitons.
Compared to interactions between identical polarized solitons \cite{Snyder-2007,Martin-2007,Zhao-2016,Zhao-2017}, orthogonally polarized solitons exhibit diverse inelastic interactions \cite{Malomed-1991,Yang-1996,Tan-2001,Rand-2007,Qin-2019,Zhao-2021}, especially in non-integrable regimes far from the Manakov case (where self-phase and cross-phase modulation coefficients are equal). 
That is why we choose circularly polarized states to study eigenvalue control.

\section{Decomposition of HOS}

\begin{figure}[htbp]
	\centering
	\includegraphics[width=86mm]{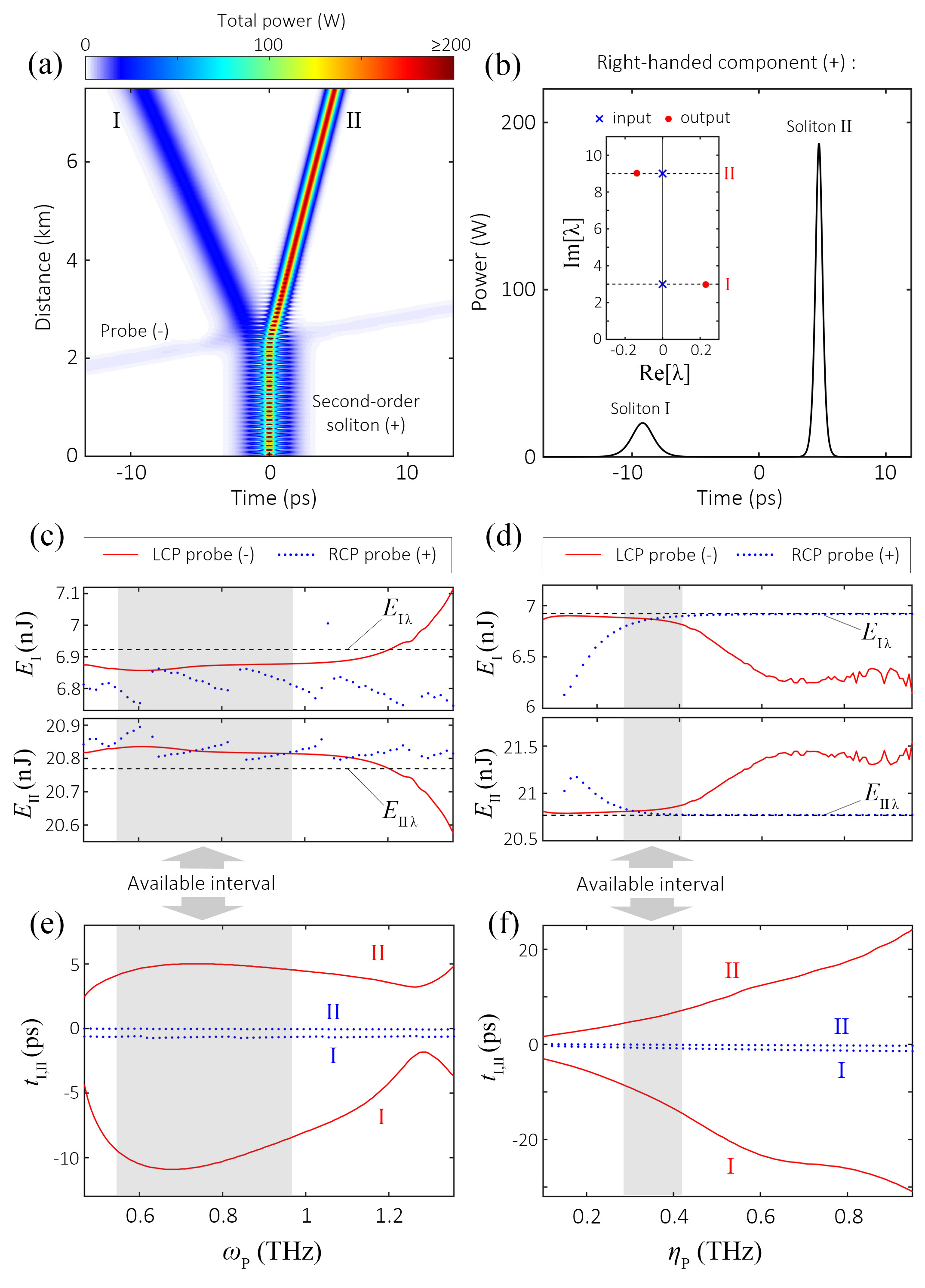}
	\caption{(a) Evolution plot of total power $|A_+|^2+|A_-|^2$ from the initial condition (\ref{eq-ini-ap}) and (\ref{eq-ini-am}), where a second-order soliton decomposes into two fundamental solitons I and II.
	(b) Distribution of power $|A_+|^2$ in the RCP component at $z=7.5\,{\rm km}$. The inset shows the discrete eigenvalues of input (blue times signs) and output (red circles) lights.
	(c) Dependence of soliton I's energy $E_{\rm I}$ (upper) and soliton II's energy $E_{\rm II}$ (lower) on the probe's frequency $\omega_{\rm p}$. 
    The gray regions are the available interval where discrete eigenvalues can be effectively measured under the condition (\ref{eq-condition}). 
    The horizontal dashed lines are the calculated energy $E_{\rm I \lambda}$ or $E_{\rm II \lambda}$ from the HOS's discrete eigenvalues.
	(d) Same as (c) except for the horizontal variable $\eta_{\rm p}$.
        (e) Dependence of center time $t_{\rm I,II}$ of soliton I and II on the probe's frequency $\omega_{\rm p}$.
        (f) Same as (e) except for the horizontal variable $\eta_{\rm p}$.
	The parameters are $N=2$, $\eta=0.9\,{\rm THz}$, $\eta_{\rm p}=0.3\,{\rm THz}$, $\omega_{\rm p}=0.9\,{\rm THz}$, and $t_{\rm p}=-4.52\,{\rm ps}$.
	}
	\label{pic-dp2}
\end{figure}

In the model (\ref{eq-model-cp}), the decomposition process of a HOS can be simulated by the following settings of parameters and input light.
We consider the experimental parameters in Ref. \cite{Kraych-2019}.
The input light has the center wavelength $1550\,{\rm nm}$ corresponding to the angular frequency $\omega=1220\,{\rm THz}$.
The group-velocity dispersion coefficient is $\beta_2=-22\,{\rm ps^2/km}$, and the Kerr nonlinear coefficient is $\sigma=1.3\,{\rm W^{-1}km^{-1}}$.
For the RCP component ($+$), the input light is a HOS with order $N$,
\begin{align}\label{eq-ini-ap}
	&A_+(t,z=0)=\sqrt{P_0}T_0\,N\eta{\rm sech}(\eta t),
\end{align}
where $\eta$ is the steepness (with the same unit as frequency).
$P_0$ and $T_0$ are respectively the reference values of optical power and wave-packet width, and here they are set as $P_0=2.3\,{\rm W}$ and $T_0=3.3\,{\rm ps}$.
For the LCP component ($-$), the input light is a fundamental soliton,
\begin{align}\label{eq-ini-am}
	&A_-(t,z=0)=\sqrt{P_0}T_0\,\eta_{\rm p}{\rm sech}[\eta_{\rm p} (t-t_{\rm p})]e^{i\omega_{\rm p}t},
\end{align}
which is called "probe soliton" here.
$t_{\rm p}$, $\omega_{\rm p}$, and $\eta_{\rm p}$ denote the time offset, relative frequency, and steepness of the soliton, respectively.
Setting $N=2$, $\eta=0.905\,{\rm THz}$, $\eta_{\rm p}=0.302\,{\rm THz}$, $\omega_{\rm p}=0.905\,{\rm THz}$, and $t_{\rm p}=-4.523\,{\rm ps}$, the model (\ref{eq-model-cp}) is numerically simulated by the split-step Fourier method \cite{Yang-book}.
The result of numerical evolution is shown in Fig. \ref{pic-dp2} (a).
The probe soliton in the LCP component collides with the second-order soliton in the RCP component at approximately $z=2.3\,{\rm km}$. 
Subsequently, the second-order soliton splits into two fundamental solitons with distinct velocities, which we label as I (left) and II (right). 
Fig. \ref{pic-dp2} (b) shows the power cross-sectional profile of the final state at $z=7.5\, {\rm km}$, revealing that the maximal power of soliton I is significantly lower than that of soliton II.

To verify whether the eigenvalues of after-collision solitons coincide with those of the original second-order soliton, we perform the following eigenvalue analysis:
When the amplitude of LCP component ($-$) is much smaller than that of RCP ($+$), the model (\ref{eq-model-cp}) can be reduced to a nonlinear Schr\"{o}dinger equation (NLSE),
\begin{align}\label{eq-nlse-cp}
	&i\partial_zA_+-\frac{\beta_2}{2}\partial_t^2A_++\frac{2}{3}\sigma|A_+|^2A_+=0.
\end{align}
For the abnormal dispersion ($\beta_2<0$), one can take $A_+=\sqrt{P_0}\psi$, $t=T_0\tau$, and $z=L_0\xi$ to non-dimensionalize the equation and obtain
\begin{align}\label{eq-nlse}
	&i\partial_\xi \psi+\frac{1}{2}\partial_\tau^2\psi+|\psi|^2\psi=0.
\end{align}
The relations between parameters are $|\beta_2|L_0=T_0^2$ and $\frac{2}{3}\sigma L_0 P_0=1$.
The reference value of distance is set as $L_0=0.5\,{\rm km}$, and thus the unit power and the unit time are separately $P_0=2.3\,{\rm W}$ and $T_0=3.3\,{\rm ps}$, as stated above.
The Lax equation of NLSE (\ref{eq-nlse}) is 
\begin{equation}
	\label{eq-lax}
	\left\{
	\begin{aligned}
		&{\boldsymbol \Phi}_\tau={\mathbf U}{\boldsymbol \Phi}, \\
		&{\boldsymbol \Phi}_\xi={\mathbf V}{\boldsymbol \Phi}, \\
	\end{aligned}
	\right.
\end{equation}
where
\begin{align}
	&{\boldsymbol \Phi}=\begin{bmatrix}\phi_1\\ \phi_2\end{bmatrix},\quad
	{\mathbf U}=\frac{\lambda}{2}
	\begin{bmatrix}-i & 0\\ 0 & i\end{bmatrix}
	+\begin{bmatrix}0 & \psi\\ -\psi^* & 0\end{bmatrix},\nonumber\\
	&{\mathbf V}=\frac{\lambda^2}{4}\begin{bmatrix}-i & 0\\ 0 & i\end{bmatrix}+\frac{\lambda}{2}\begin{bmatrix}0 & \psi\\ -\psi^* & 0\end{bmatrix}+\frac{1}{2}\begin{bmatrix}i|\psi|^2 & i\psi_t\\ i\psi^* & -i|\psi|^2\end{bmatrix}.
\end{align}
$\lambda$ is the Zakharov-Shabat eigenvalue.
Therefore, given the temporal profile of a wavefunction, we can calculate its corresponding eigenvalues using the first equation of the Lax pair (\ref{eq-lax}). 
%Crucially, the Lax equations ensure that these eigenvalues remain invariant at all propagation distances. 
Recognizing that obtaining eigenvalues analytically is generally difficult, we will employ the Fourier collocation method to numerically solve for the eigenvalues \cite{Yang-book}.

For a fundamental soliton, the real part of $\lambda$ determines its velocity, while the imaginary part determines its steepness and thus is a key quantity to distinguish solitons.
In the case of Fig. \ref{pic-dp2} (a), the discrete eigenvalues $\lambda$ of input and output lights in the RCP component (+) are shown in the inset of Fig. \ref{pic-dp2} (b).
One can see that the discrete eigenvalues of input and output have different real parts but almost the same imaginary parts, i.e., ${\rm Im}[\lambda]=3$ and $9$.
Therefore, the splitting process can be regarded as an eigenvalue splitting, as the imaginary part of eigenvalues remain unchanged.
Considering that this process involves the splitting of a composite structure into multiple fundamental structures, we here term this phenomenon the decomposition of HOSs.

Next, we investigate the influence of the initial configuration of the probe soliton on the accuracy of after-collision eigenvalue measurements, as well as compare the decomposition efficiency with that of the single-component scenario as discussed in Ref. \cite{Hasegawa-1993}. 
In Fig. \ref{pic-dp2} (c), the red solid curves illustrate the dependence of the pulse energy ($E_{\rm I}$ and $E_{\rm II}$) of the after-collision solitons on the relative frequency $\omega_{\rm p}$ and steepness $\eta_{\rm p}$ of the probe soliton. 
Here, the pulse energy is calculated by temporal integration of the power near the peak. 
The upper and lower panels correspond to the results for solitons I and II, respectively. 
For reference, the ideal energy values ($E_{\rm I\lambda}$ and $E_{\rm II\lambda}$) of fundamental solitons which has the discrete eigenvalues of the original HOS are marked with black dashed curves (calculated by $E_{\rm I\lambda}=2P_0T_0{\rm Im}[\lambda_{\rm I}]$ and $E_{\rm II\lambda}=2P_0T_0{\rm Im}[\lambda_{\rm II}]$).
It is observed that when the relative frequency is not excessively large, the pulse energies of the two decomposed fundamental solitons closely match the ideal values. 
Furthermore, compared with Ref. \cite{Hasegawa-1993}, blue circular markers represent measurement results under conditions where the probe soliton is also RCP, while maintaining identical incident waveform parameters. 
In this case, the energy distribution of the probe soliton significantly perturbs the energy measurements of the after-collision solitons, resulting in periodic serrated structures in the energy profiles. 
Compared to the scenario with a LCP probe soliton, the measurements by a RCP probe exhibit larger deviations from the ideal energy values across most parameter ranges.
Fig. \ref{pic-dp2} (d) shows the dependence of $E_{\rm I}$ and $E_{\rm II}$ on $\eta_{\rm p}$.
When $\eta_{\rm p}$ is small, the energy of the after-collision solitons closely aligns with the ideal energy values, indicating that the process constitutes an effective decomposition. 
As $\eta_{\rm p}$ increases, the deviation between the measured and ideal energies grows progressively. 
In contrast, when the probe soliton shares the same RCP state as the HOS, the decomposition accuracy improves at larger $\eta_{\rm p}$ values, a trend opposite to that observed with a LCP probe soliton.

The influence of the relative frequency $\omega_{\rm p}$ and steepness $\eta_{\rm p}$ of the probe soliton on the time coordinate of the after-collision solitons is shown in Figs. \ref{pic-dp2} (e) and (f). 
These results can demonstrate the efficiency of HOS decomposition, because a larger temporal separation of after-collision solitons enables the discrete eigenvalues to be resolved at shorter propagation distances.
One can find that a modest $\omega_{\rm p}$ or a relatively large $\eta_{\rm p}$ can improve the efficiency of HOS decomposition.
To ensure both accuracy in soliton energy measurement and efficiency in decomposition, we show optimal ranges of $\omega_{\rm p}$ and $\eta_{\rm p}$ (see the gray regions in Fig. \ref{pic-dp2}), whose conditions are
\begin{align}\label{eq-condition}
	&|E_{\rm I}-E_{\rm I\lambda}|<0.1\,{\rm n\,J},\;
	|E_{\rm II}-E_{\rm II\lambda}|<0.1\,{\rm n\,J},\;
	|t_{\rm I}-t_{\rm II}|>2\,{\rm ps}.
\end{align}
In contrast, when the probe soliton shares the same RCP state as the HOS, the two after-collision solitons have insufficient temporal separation, and thus one cannot effectively measure eigenvalues. 
These observations reveal that the decomposition efficiency induced by a LCP probe soliton exceeds that induced by a RCP probe soliton by more than a factor of 15.

\begin{figure}[htbp]
  \centering
  \includegraphics[width=86mm]{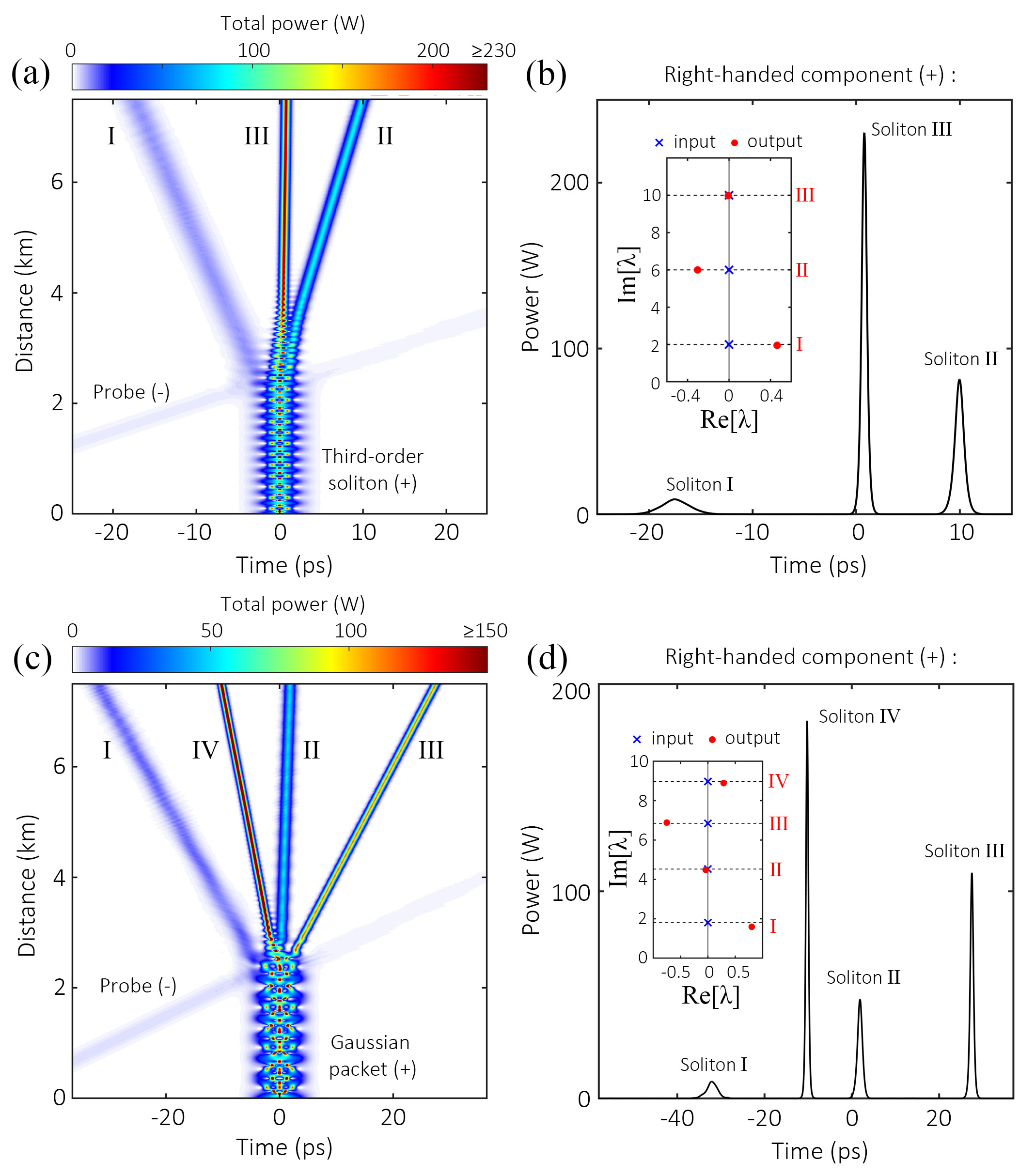}
  \caption{(a) Evolution plot of total power $|A_+|^2+|A_-|^2$ from the initial condition (\ref{eq-ini-ap}) and (\ref{eq-ini-am}), where a second-order soliton decomposes into two fundamental solitons I and II.
  (b) Distribution of power $|A_+|^2$ in the RCP component at $z=7.5\,{\rm km}$. The inset shows the discrete eigenvalues of input (blue times signs) and output (red circles) lights.
  The parameters are $N=3$, $\eta=0.603\,{\rm THz}$, $\eta_{\rm p}=0.302\,{\rm THz}$, $\omega_{\rm p}=0.905\,{\rm THz}$, and $t_{\rm p}=-4.523\,{\rm ps}$.
  (c) and (d) are the same as (a) and (b) except that the input light of component ($+$) is a Gaussian packet (\ref{eq-ini-gauss}), and the parameters are $P=57.7\,{\rm W}$ and $w=3.3\,{\rm ps}$.}
  \label{pic-dp3}
\end{figure}

An interesting phenomenon can be observed in Figs. \ref{pic-dp2} (d) and (f): within a certain distance, one cannot simultaneously measure the soliton eigenvalues (i.e., pulse energy) and their temporal coordinates with high accuracy. 
To measure soliton eigenvalues as precisely as possible, we need to reduce the steepness $\eta_{\rm p}$. 
However, this causes the after-collision solitons to be too close in time, making it difficult to accurately determine their temporal coordinates. 
Conversely, achieving sufficient temporal separation between solitons requires a larger $\eta_{\rm p}$, but this compromises the accuracy of eigenvalue measurements. 
This behavior resembles the uncertainty principle in quantum mechanics. 
By analogy, we try to give an uncertainty relationship for soliton eigenvalue measurements: $\Delta E \Delta\, t\geq C$, where $\Delta E$, $\Delta\, t$, and $C$ are the levels of pulse energy, temporal coordinate, and an unknown quantity. 
The deeper physical mechanisms behind this phenomenon require further investigation.

We also use the method of collision between circularly polarized lights to decompose higher-order solitons. 
The condition of input light is still Eqs. (\ref{eq-ini-ap}) and (\ref{eq-ini-am}), and Figs. \ref{pic-dp3} (a) and (b) show the decomposition result of a third-order soliton. 
After splitting, the imaginary parts of the eigenvalues of the three fundamental solitons exactly match those of the discrete eigenvalues forming the original third-order soliton. 
Using the same method, we also decompose a Gaussian wave packet.
Holding the LCP component ($-$) unchanged, the RCP component ($+$) is set as a Gaussian packet,
\begin{align}\label{eq-ini-gauss}
	&A_+(t,z=0)=\sqrt{P}e^{-t^2/(2w^2)},
\end{align}
where $w$ is its width.
The result of its decomposition is shown in Figs. \ref{pic-dp3} (c) and (d). 
It is found that after collision four separate fundamental solitons appear, the imaginary parts of whose eigenvalues also perfectly align with their discrete eigenvalues, suggesting it is essentially a fourth-order soliton.

\section{Reconstruction of HOS}

\begin{figure}[htbp]
\centering
\includegraphics[width=86mm]{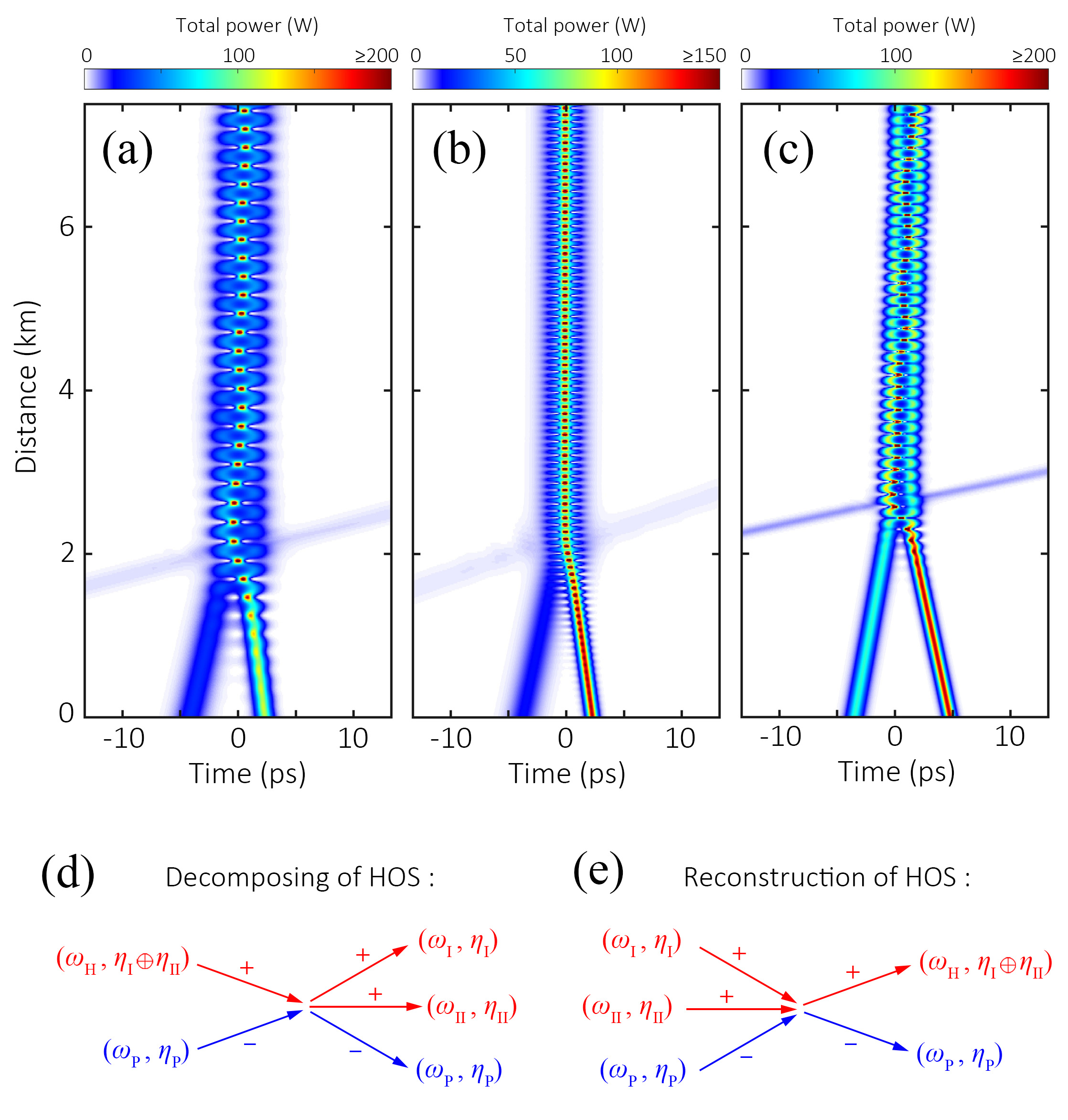}
\caption{(a-c) Evolution plot of total power $|A_+|^2+|A_-|^2$ from the initial condition (\ref{eq-inirc-ap}) and (\ref{eq-inirc-am}), where a second-order soliton decomposes into two fundamental solitons I and II. 
(a), (b), and (c) correspond to discrete eigenvalue combinations of $({\rm Im}[\lambda_{\rm I}],{\rm Im}[\lambda_{\rm II}])=$(3, 6), (3, 9), and (6, 9), respectively. 
(d) Decomposing and (e) reconstruction processes of HOS shown in a form similar to Feynman diagram, which provide possibility to make all-optical demultiplexer and multiplexer, respectively.}
  \label{pic-rc}
\end{figure}

Through the interaction between circularly polarized lights, soliton eigenvalues can also recombine, manifested as the reconstruction of multiple fundamental solitons into a HOS. 
This reconstruction is achieved by colliding two RCP fundamental solitons and one LCP fundamental soliton at a certain position.
For the RCP component ($+$), the input light is composed of two fundamental solitons,
\begin{align}\label{eq-inirc-ap}
	A_+(t,z=0)=\sqrt{P_0}&T_0\big[\eta_{\rm I}{\rm sech}[\eta_{\rm I} (t-t_{\rm I})]e^{i\omega_{\rm I}(t-t_{\rm I})}\nonumber\\
	&+\eta_{\rm II}{\rm sech}[\eta_{\rm II} (t-t_{\rm II})]e^{i\omega_{\rm II}(t-t_{\rm II})}\big],
\end{align}
where $\omega_{\rm I,II}$, $\eta_{\rm I,II}$, and $t_{\rm I,II}$ are their relative frequency, steepness, and temporal coordinates, respectively.
For the LCP component ($-$), the input light is a fundamental soliton,
\begin{align}\label{eq-inirc-am}
	&A_-(t,z=0)=\sqrt{P_0}T_0\,\eta_{\rm p}{\rm sech}[\eta_{\rm p} (t-t_{\rm p})]e^{i\omega_{\rm p} t},
\end{align}
where $\omega_{\rm p}$, $\eta_{\rm p}$, and $t_{\rm p}$ are its relative frequency, steepness, and temporal coordinates, respectively.
Figs. \ref{pic-rc} (a), (b), and (c) demonstrate the examples of reconstructing second-order solitons using different eigenvalue combinations by fine-tuning the inverse process of second-order soliton decomposition. 
Panels (a), (b), and (c) correspond to discrete eigenvalue combinations of $({\rm Im}[\lambda_{\rm I}],{\rm Im}[\lambda_{\rm II}])=$(3, 6), (3, 9), and (6, 9), respectively. 
The parameter settings are as follows:
In panel (a), we have $\eta_{\rm I}=0.905\,{\rm THz}$, $\eta_{\rm II}=1.81\,{\rm THz}$, $\eta_{\rm p}=0.302\,{\rm THz}$, $\omega_{\rm I}=0.075\,{\rm THz}$, $\omega_{\rm II}=-0.036\,{\rm THz}$, $\omega_{\rm p}=1.206\,{\rm THz}$, $t_{\rm I}=-4.146\,{\rm ps}$, $t_{\rm II}=2.322\,{\rm ps}$, and $t_{\rm p}=-61.026\,{\rm ps}$;
in panel (b), we have $\eta_{\rm I}=0.905\,{\rm THz}$, $\eta_{\rm II}=2.714\,{\rm THz}$, $\eta_{\rm p}=0.302\,{\rm THz}$, $\omega_{\rm I}=0.069\,{\rm THz}$, $\omega_{\rm II}=-0.042\,{\rm THz}$, $\omega_{\rm p}=0.905\,{\rm THz}$, $t_{\rm I}=-3.768\,{\rm ps}$, $t_{\rm II}=2.283\,{\rm ps}$, and $t_{\rm p}=-44.443\,{\rm ps}$;
in panel (c), we have $\eta_{\rm I}=1.81\,{\rm THz}$, $\eta_{\rm II}=2.714\,{\rm THz}$, $\eta_{\rm p}=0.603\,{\rm THz}$, $\omega_{\rm I}=0.583\,{\rm THz}$, $\omega_{\rm II}=-0.068\,{\rm THz}$, $\omega_{\rm p}=1.494\,{\rm THz}$, $t_{\rm I}=-3.516\,{\rm ps}$, $t_{\rm II}=4.776\,{\rm ps}$, and $t_{\rm p}=-87.459\,{\rm ps}$.
These results suggest that such reconstruction process of a HOS is physically feasible, and thus provides a possible all-optical scheme of eigenvalue-based multiplexers, which is significantly different from the mathematical tools, such as nonlinear Fourier transformation and Darbox transformation.
Beyond the above examples, more HOSs with various eigenvalue combinations could be reconstructed.

Finally, we compare the decomposition/reconstruction process of HOSs with inelastic light scattering phenomena in nonlinear optics, discussing their differences and connections. 
Here, we treat solitons as "photon-like particle" characterized by two fundamental parameters: frequency and steepness. 
Specifically, $\omega_{\rm H}$, $\omega_{\rm p}$, $\omega_{\rm I}$, and $\omega_{\rm II}$ denote the frequencies of the HOS, probe soliton, and the two split fundamental solitons, respectively, while $\eta_{\rm I}\oplus\eta_{\rm II}$, $\eta_{\rm p}$, $\eta_{\rm I}$, and $\eta_{\rm II}$ represent their corresponding steepness values.
Schematics of the decomposition and reconstruction processes of a high-order soliton, drawn in a style similar to Feynman diagrams, are shown in Figs. \ref{pic-rc} (d) and (e). 
Steepness serves as an important quantity for solitons: within the same system, steepness remains invariant and cannot be further subdivided, which is similar to a particle’s charge or spin. 
However, unlike these intrinsic properties, steepness exhibits unique property of recombination and decomposition, as demonstrated in this work. 
We expect that these intriguing characteristics may make steepness a key parameter (potentially replacing or complementing frequency, charge, and other conventional quantities) to enable novel physical phenomena.

\section{Conclusion}

In summary, we develop an effective method for controlling soliton eigenvalues in optical fibers through interactions between circularly polarized lights. 
Using this approach, we achieve eigenvalue decomposition of HOSs and subsequently reconstruct them via the inverse decomposition process. 
Compared to interactions involving same-polarization lights, our method achieves over 15-fold improvement in decomposition/reconstruction efficiency while maintaining high measurement accuracy for eigenvalues.
We further investigate how the frequency and steepness of probe solitons affect both measurement accuracy and decomposition efficiency, identifying parameter ranges that enable effective decomposition. 
Notably, successful decomposition requires the steepness (or pulse energy) of the probe soliton to remain within a specific range: high values reduce measurement accuracy, while low values reduce decomposition efficiency. 
This has similarity to the uncertainty principle in quantum mechanics.
Finally, we analyze the important role of soliton steepness in eigenvalue control and its connection to fundamental physical properties.

\section*{Acknowledgement}
The authors thank Prof. Li-Chen Zhao and Dr. Liang Duan for their helpful discussions.
This work was supported by the Fundamental Research Funds for the Central Universities (20103258347); Scientific Research Innovation Capability Support Project for Young Faculty (ZYGXQNJSKYCXNLZCXM-123); the National Natural Science Foundation of China (62335018); the National Key Research and Development Program of China (2021YFF0700303); Key Research and Development Program of Shaanxi Province (2024GH-ZDXM-05); the Natural Science Foundation of Shaanxi Province (2025JC-YBQN-819, 2025JC-YBMS-695); China Postdoctoral Science Foundation (2024M762528); Xidian University Specially Funded Project for Interdisciplinary Exploration (TZJH2024040, TZJH2024044); the Fundamental Research Funds for the Central Universities (ZYTS25127, ZYTS25133); the Key Scientific Research Program of Shaanxi Provincial Department of Education (24JR065).

%\appendix
%\setcounter{equation}{0}
%\renewcommand\theequation{A\arabic{equation}}

%\section*{Appendix I: Propagation equation of light in the Frenet coordinate system}
%\label{app2}

\end{document}